\documentclass[a4paper,11pt]{article}
\usepackage{pos}

\title{Probing the interior of Earth using oscillating neutrinos at INO-ICAL}

\author*[a,b,c]{Anil Kumar}
\author[a,d]{Anuj Kumar Upadhyay}
\author[a,c,e]{Sanjib Kumar Agarwalla}
\author[f]{Amol Dighe}

\affiliation[a]{Institute of Physics, Sachivalaya Marg, Sainik School Post, Bhubaneswar 751005, India}

\affiliation[b]{Applied Nuclear Physics Division, Saha Institute of
	Nuclear Physics, Block AF, Sector 1, Bidhannagar, Kolkata 700064, India}

\affiliation[c]{Homi Bhabha National Institute, Anushakti Nagar, Mumbai 400094, India}

\affiliation[d]{Department of Physics, Aligarh Muslim University, Aligarh 202002, India}

\affiliation[e]{Department of Physics and Wisconsin IceCube Particle Astrophysics Center, University of Wisconsin, Madison, Wisconsin 53706, USA}

\affiliation[f]{Tata Institute of Fundamental Research, Homi Bhabha Road, Colaba, Mumbai 400005, India}

\emailAdd{anil.kumar@desy.de}
\emailAdd{anuju@iopb.res.in}
\emailAdd{sanjib@iopb.res.in}
\emailAdd{amol@theory.tifr.res.in}

\abstract{Atmospheric neutrinos offer the possibility of exploring the internal structure of Earth. This information is complementary to the traditional probes of seismic and gravitational studies. While propagating through Earth, the multi-GeV neutrinos encounter the Earth’s matter effects due to the coherent forward scattering with the ambient electrons, which alters the neutrino oscillation probabilities. We present how well an atmospheric neutrino oscillation experiment like the 50 kt Iron Calorimeter (ICAL) detector at India-based Neutrino Observatory would validate the presence of Earth’s core, measure the location of the core-mantle boundary (CMB), and probe the dark matter (DM) inside the Earth in a unique way through Earth matter effects in neutrino oscillations. Owing to good angular resolution, ICAL can observe the core-passing neutrinos efficiently. Due to its magnetized setup, it would be able to observe neutrinos and antineutrinos separately. With 500 kt$\cdot$yr exposure, the presence of Earth’s core can be independently confirmed at ICAL with a median $\Delta\chi^2$ of 7.45 (4.83) for normal (inverted) mass ordering. With 1000 kt$\cdot$yr exposure, ICAL would be able to locate the CMB with a precision of about $\pm 250$ km at $1\sigma$. It would also be sensitive to the possible presence of dark matter with 3.5\% of the mass of Earth at $1\sigma$. The charge identification capability of ICAL would play an important role in achieving these precisions.}

\FullConference{%
	*** The European Physical Society Conference on High Energy Physics (EPS-HEP 2023), ***\\
	*** 21-25 August 2023 ***\\
	*** Hamburg, Germany ***
}

\begin{document}
\maketitle

\section{Introduction and Motivation}
\label{intro}

The information about the interior of Earth has primarily been obtained using traditional indirect probes in seismic and gravitational studies. The seismic wave propagation data has been used to infer the most widely accepted density model of Earth, the Preliminary Reference Earth Model (PREM)~\cite{Dziewonski:1981xy}. The complementary information about the internal structure of Earth may also be obtained by observing the upward-going atmospheric neutrinos that pass through Earth. While traversing through the different regions inside Earth, multi-GeV atmospheric neutrinos encounter the Earth's matter effects~\cite{Wolfenstein:1977ue} due to the coherent forward scattering with the ambient electrons. These matter effects depend upon the energy of neutrinos and the number density of electrons in the medium. Hence, the probability patterns for neutrino flavor conversion depend upon the abrupt changes in density, such as the transition from Earth's core to mantle~\cite{Kumar:2021faw}, and the location of the core-mantle boundary (CMB)~\cite{Upadhyay:2022jfd}. Furthermore, the neutrino oscillation dependence on electron number density can also be interpreted in terms of the baryonic matter density inside Earth. A deficit in the observed amount of baryonic mass as compared to the prediction from the gravitational measurements may be interpreted as a possible presence of dark matter (DM) inside Earth~\cite{Upadhyay:2021kzf}.

In these studies, we explore the effects of the presence of Earth's core~\cite{Kumar:2021faw}, modification of CMB location~\cite{Upadhyay:2022jfd}, and the possible presence of DM~\cite{Upadhyay:2021kzf}, on the neutrino oscillation patterns using a proposed 50 kt Iron Calorimeter (ICAL) detector at the India-based Neutrino Observatory (INO)~\cite{ICAL:2015stm}. Owing to the charge identification (CID) capability, ICAL would be able to detect atmospheric neutrinos and antineutrinos separately in a multi-GeV energy range over a wide range of baselines. About 20\% of upward-going neutrinos observed at ICAL, pass through the core of Earth. These unique features make ICAL an excellent tool for probing Earth's core using atmospheric neutrinos.

\section{Testing Alternative Density Profiles of Earth}
\label{various_profile}

\begin{figure}[h!]
	\centering
	\includegraphics[width=0.49\linewidth]{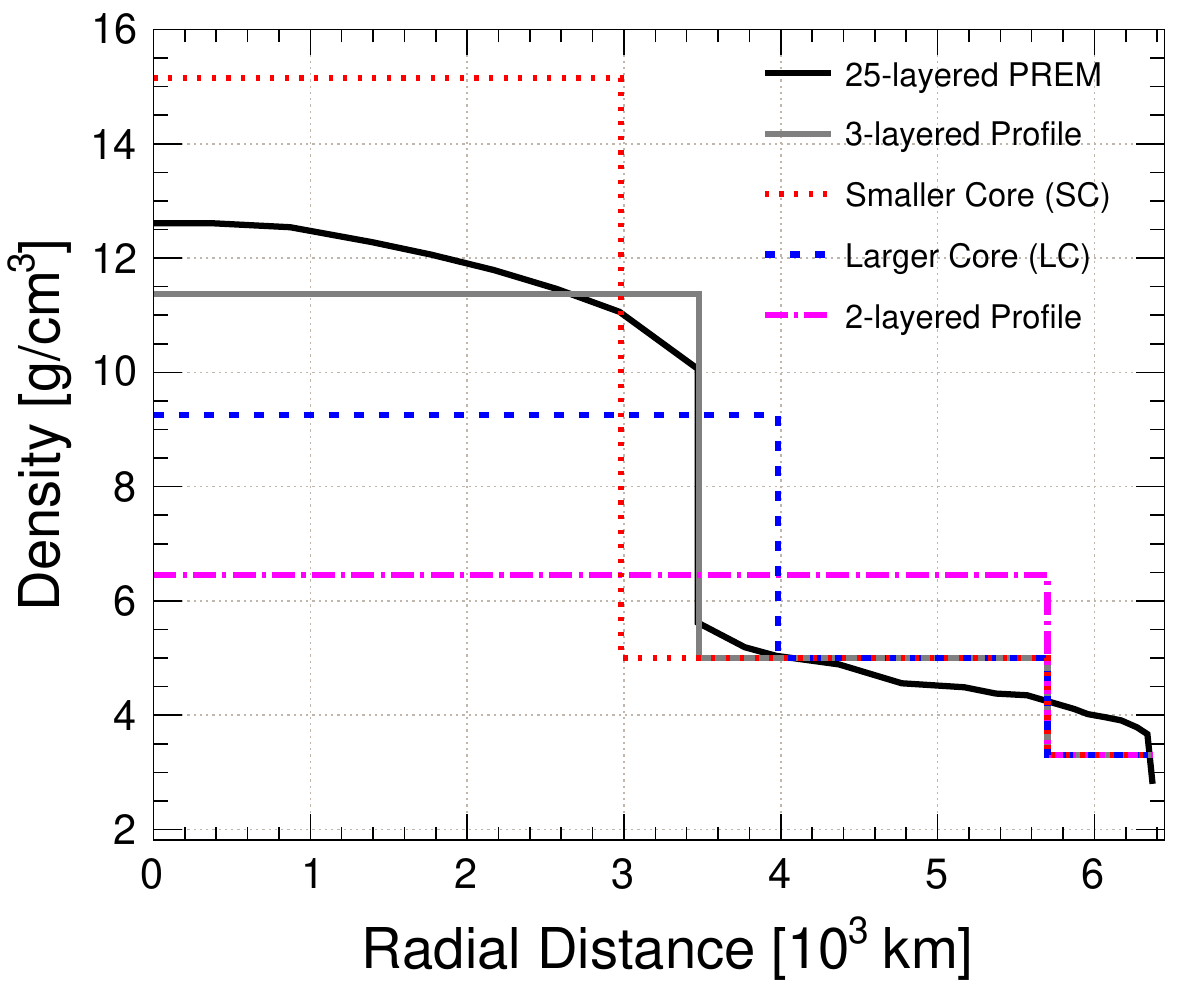}
	\includegraphics[width=0.49\linewidth]{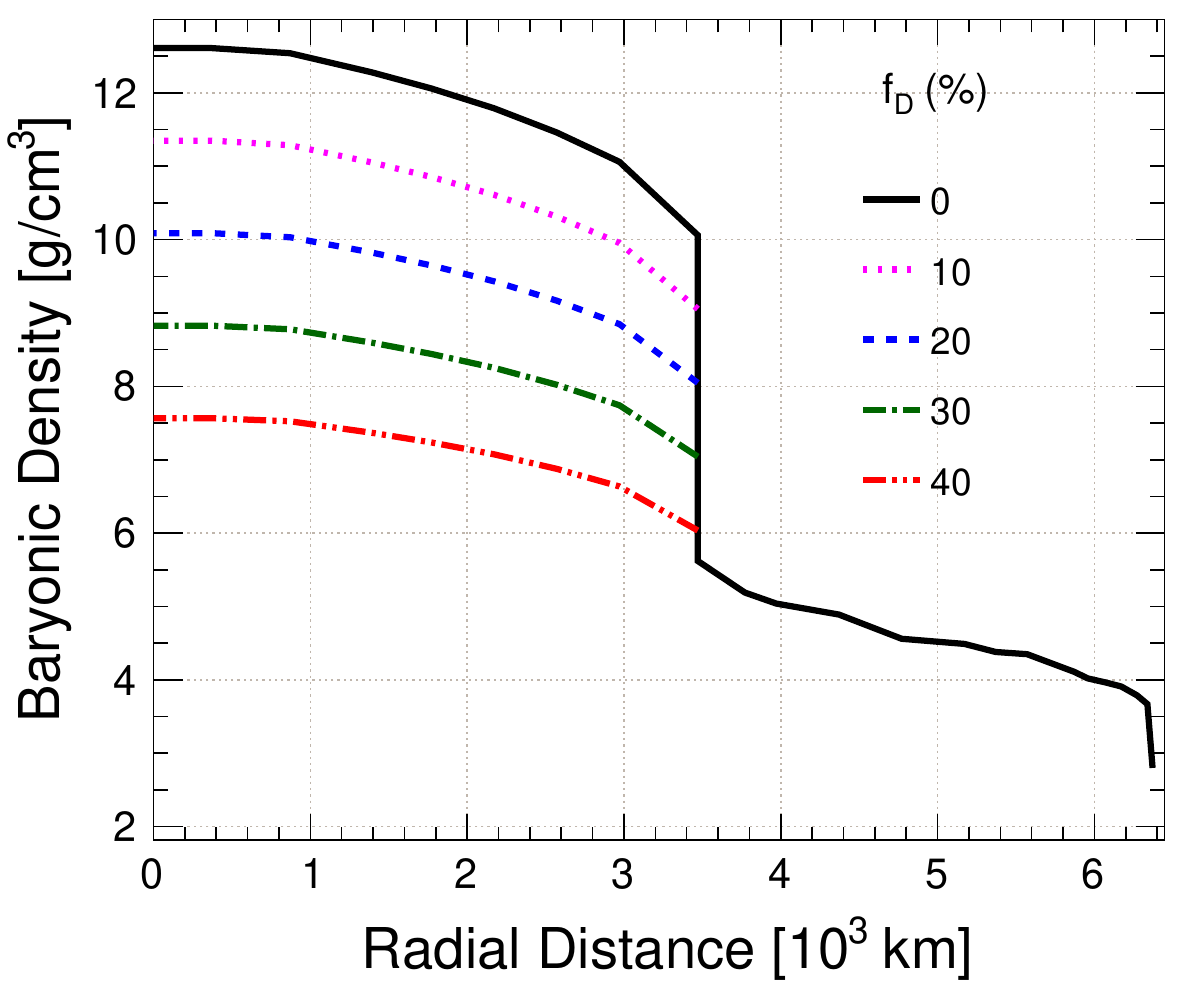}
	\caption{Left: Alternative density distribution profiles of Earth as a function of radial distance~\cite{Kumar:2021faw,Upadhyay:2022jfd}. Right: Radial density distribution of baryonic matter inside Earth guided by 25-layered PREM profile for some representative choices of uniform DM fraction $f_\text{D}$ inside the core. The right figure is taken from the Ref.~\cite{Upadhyay:2021kzf}.}
	\label{fig:profiles}
\end{figure}

In this work, we consider the alternative density profiles of Earth with different numbers of layers. In Fig.~\ref{fig:profiles}, the solid black curves in both the left and right panels represent the PREM profile with 25 layers. In the left panel, the solid grey curve represents a simplified three-layered profile of Earth consisting of a core, inner mantle, and outer mantle, whereas dotted-dashed magenta curve depicts a two-layered profile consisting of inner and outer mantle but no core. By distinguishing this two-layered profile with respect to the three-layered profile using atmospheric neutrinos at ICAL, we can validate the presence of Earth' core~\cite{Kumar:2021faw}. The dotted-red and dashed-blue curves in the left panel correspond to the three-layered profile with a modified location of CMB radius by $\Delta R_\text{CMB} = -500$ km (smaller core or SC) and $\Delta R_\text{CMB} = +500$ km (larger core or LC), respectively, with respect to the standard $R_\text{CMB} = 3480$ km. Note that the mass of core and the total mass of Earth remain constant during these CMB modifications and this scenario is referred to as ``Case-II'' in Ref.~\cite{Upadhyay:2022jfd}. The right panel of Fig.~\ref{fig:profiles} illustrates different baryonic density profiles that incorporate various choices of DM fraction ($f_\text{D}$) within the Earth's core~\cite{Upadhyay:2021kzf}. The baryonic density of the core is decreased by a uniform DM fraction $f_\text{D}$ to compensate for the possible mass of DM inside core ensuring that the total mass of Earth remains invariant. 

\section{Results}
\label{result}

The unoscillated neutrino events with the ICAL geometry are simulated using the NUANCE Monte Carlo neutrino event generator with the Honda 3D atmospheric neutrino flux at the INO location~\cite{ICAL:2015stm}. The oscillation probabilities and detector properties are incorporated as described in Refs.~\cite{Kumar:2021faw,Upadhyay:2021kzf,Upadhyay:2022jfd}. The expected median sensitivity for ICAL is calculated following the procedure described in Refs.~\cite{Kumar:2021faw,Upadhyay:2021kzf,Upadhyay:2022jfd}. The optimized binning scheme for validating the presence of the core inside Earth is taken from Table 5 in Ref.~\cite{Kumar:2021faw}, whereas the optimized binning scheme given in Table 4 in Ref.~\cite{Upadhyay:2022jfd} is used for locating the CMB and probing the possible presence of DM inside Earth.

\begin{figure}[h!]
	\centering
	\includegraphics[width=0.49\linewidth]{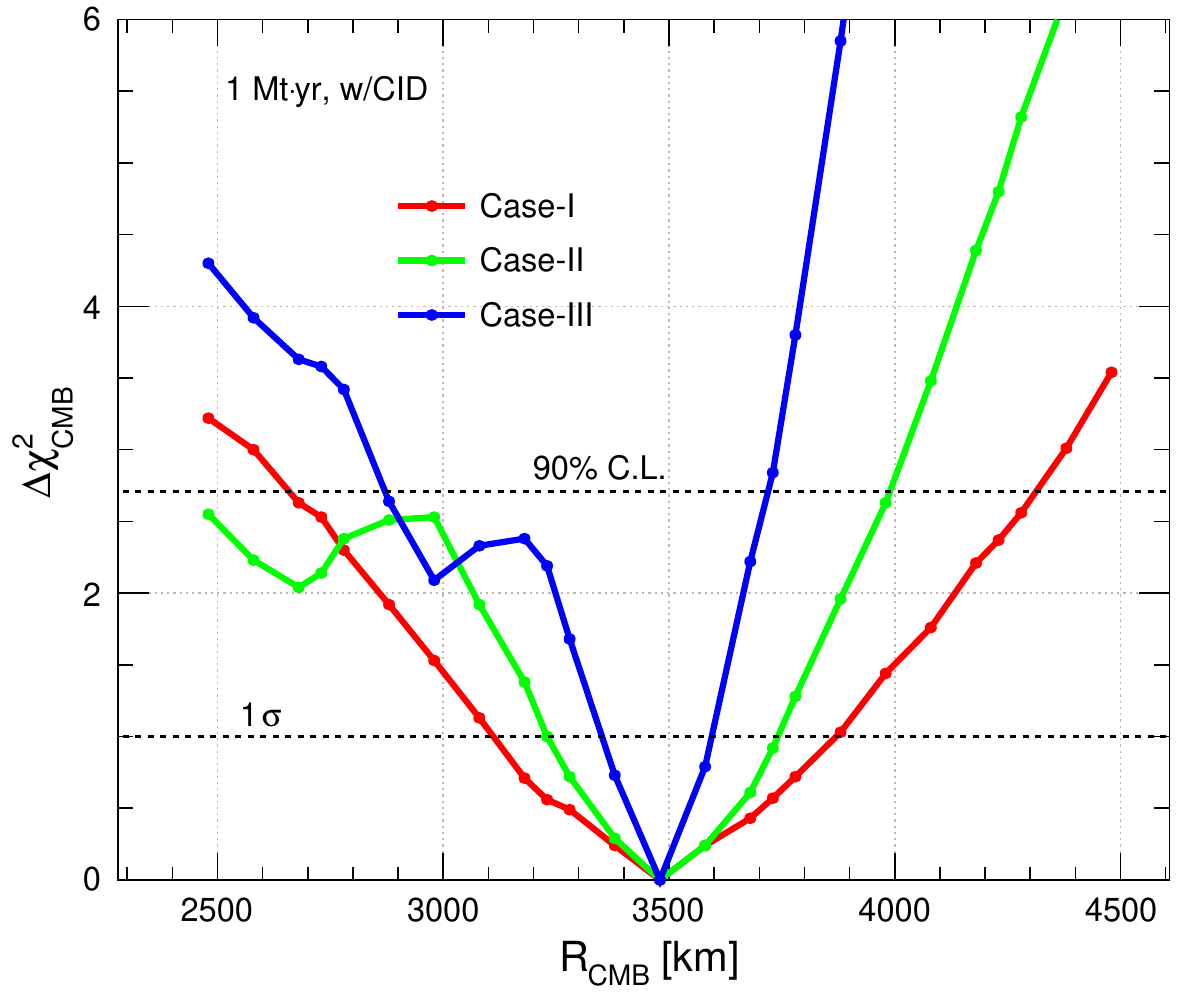}
	\includegraphics[width=0.49\linewidth]{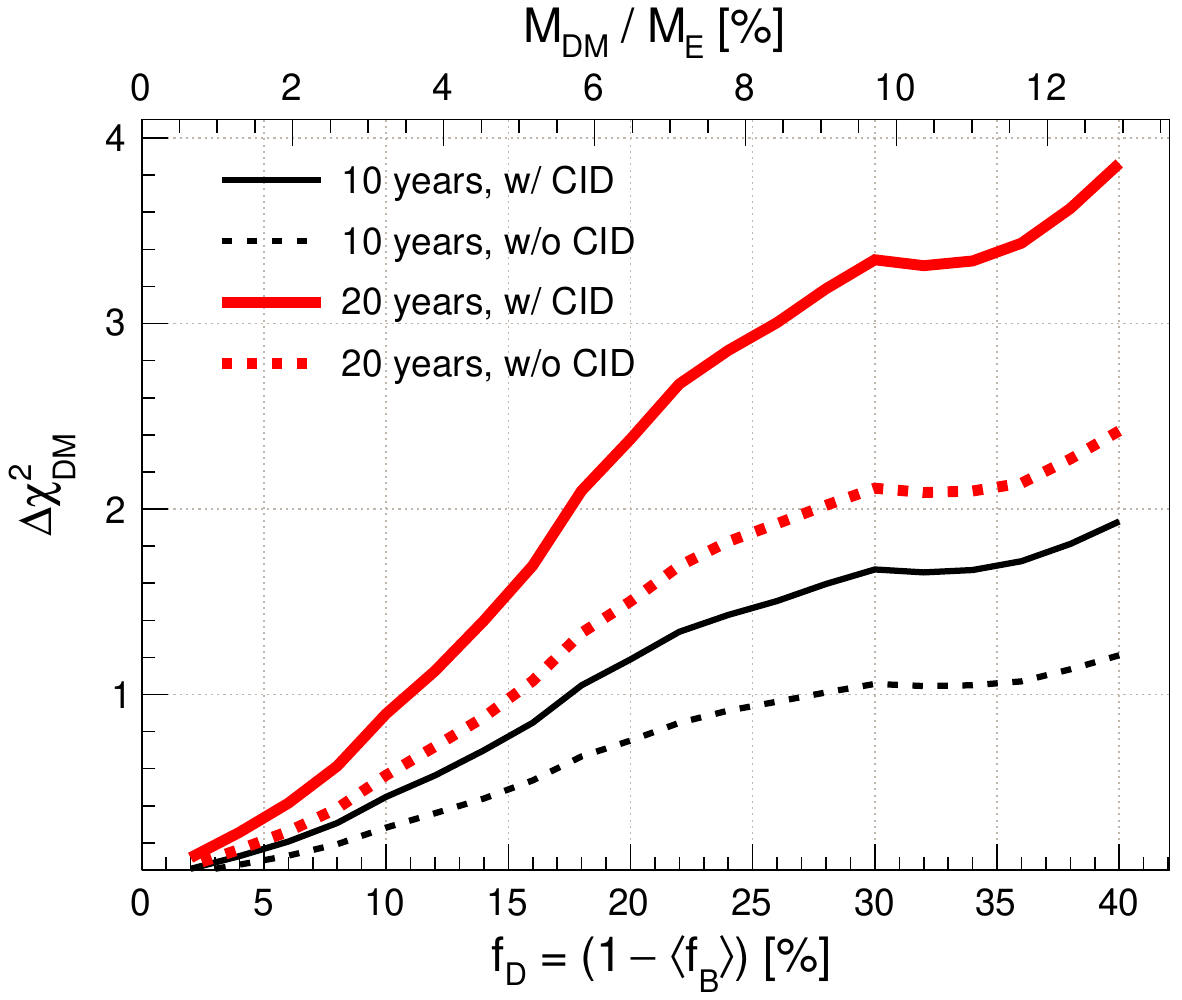}
	\caption{Left: The expected median $\Delta\chi^2_\text{CMB}$ as functions of location of the CMB $(R_\text{CMB})$. The red, green, and blue curves present the sensitivity obtain by modifying the CMB in three different ways: Case-I, Case-II, and Case-III, respectively~\cite{Upadhyay:2022jfd}. The left figure is taken from Ref.~\cite{Upadhyay:2022jfd}. Right: The expected median $\Delta\chi^2_\text{DM}$ as a function of DM fraction $f_\text{D}$. The top scale on the plot corresponds to the DM mass fraction in term of Earth's mass ($M_\text{E}$). Thin (thick) curve presents the sensitivity using 10 (20) years of exposure, whereas solid and dashed curves correspond to with and without CID capability of ICAL, respectively. The right figure is taken from Ref.~\cite{Upadhyay:2021kzf}.}
	\label{fig:results}
\end{figure}

ICAL would be able to validate the presence of a denser core inside Earth with a median sensitivity $\Delta\chi^2$ of 7.75 for normal mass ordering (NO) by rejecting the coreless profile in the fit with respect to the PREM profile in MC data with 500 kt$\cdot$yr exposure and using CID capability~\cite{Kumar:2021faw}.
The left plot of Fig.~\ref{fig:results} shows the expected median sensitivity for the location of CMB radius using the ICAL detector for three different ways of $R_\text{CMB}$ modification~\cite{Upadhyay:2022jfd}. From the plot, we can observe that ICAL would be able to locate the CMB with a precision of about $\pm250$ km at $1\sigma$ for NO using 1000 kt$\cdot$yr exposure and CID capability~\cite{Upadhyay:2022jfd}. The right plot of Fig.~\ref{fig:results} presents the expected median sensitivity for the possible presence of DM inside Earth~\cite{Upadhyay:2021kzf}. ICAL would be sensitive to about 5.5\% (3.5\%) of the mass of Earth as DM at $1\sigma$ for NO with an exposure of 10 (20) years with CID~\cite{Upadhyay:2021kzf}. 

We have also observed that the CID capability of the ICAL detector would play an essential role in achieving these sensitivities.  Without CID capability, the sensitivity for validating the presence of Earth's core would decrease to a $\Delta\chi^2$ value of 3.76~\cite{Kumar:2021faw}, and the precision in locating the CMB radius would deteriorate to around $\pm 330$ km at $1\sigma$~\cite{Upadhyay:2022jfd}. Similarly, in the absence of CID capability, ICAL would be sensitive to about $5\%$ mass of Earth as DM at $1\sigma$ with 20 years of exposure~\cite{Upadhyay:2021kzf}.

\section{Conclusions}
\label{sec:conclusion}

In these studies, we exploit the weak interactions of neutrinos to probe the internal structure of Earth in an independent and complementary way to the traditional seismic and gravitational measurements. We demonstrate that using matter effects in neutrino oscillations, an atmospheric neutrino experiment like INO-ICAL, would be able to validate the presence of a denser core inside Earth, measure the location of the CMB, and probe the possible presence of the DM inside Earth.

\acknowledgments We acknowledge financial support from the DAE, DST, DST-SERB, Govt. of India, INSA, and the USIEF. A.K.U. acknowledges financial support from the DST, Govt. of India (DST/INSPIRE Fellowship/2019/IF190755).

\setlength{\bibsep}{5pt plus 10ex}


\begin{thebibliography}{10}
	
	\bibitem{Dziewonski:1981xy}
	A.~M. Dziewonski and D.~L. Anderson, {\em Phys. Earth Planet. Interiors} {\bf 25} (1981) 297--356.
	
	\bibitem{Wolfenstein:1977ue} 
	L.~Wolfenstein,  {\em PRD} {\bf 17} (1978) 2369--2374.
	
	\bibitem{Kumar:2021faw}
	A.~Kumar and S.~K. Agarwalla,  {\em JHEP} {\bf 08} (2021) 139.
	
	\bibitem{Upadhyay:2022jfd}
	A.~K. Upadhyay, A.~Kumar, S.~K. Agarwalla, and A.~Dighe,  {\em JHEP} {\bf 04} (2023) 068.
	
	\bibitem{Upadhyay:2021kzf}
	A.~K. Upadhyay, A.~Kumar, S.~K. Agarwalla, and A.~Dighe, {\em PRD} {\bf 107} (2023), 11 115030.
	
	\bibitem{ICAL:2015stm}
	{\bf ICAL} Collaboration, S.~Ahmed et~al., {\em Pramana} {\bf 88} (2017), no.~5 79.
	
\end{thebibliography}
\end{document}